\documentclass[onecolumn,showpacs,preprintnumbers,amsmath,amssymb,superscriptaddress,prb]{revtex4}

\usepackage{graphicx}
\usepackage{dcolumn}
\usepackage{bm}

\begin{document}


\title{Current-induced switching of a single-molecule magnet\\ with arbitrary oriented easy axis }

\author{Maciej Misiorny}
\affiliation{%
Department of Physics, Adam Mickiewicz University, 61-614
Pozna\'{n}, Poland
}%
\author{J\'{o}zef Barna\'{s}}%
 \email{barnas@amu.edu.pl}
\affiliation{%
Department of Physics, Adam Mickiewicz University, 61-614
Pozna\'{n}, Poland
}%
\affiliation{%
Institute of Molecular Physics, Polish Academy of Sciences, 60-179
Pozna\'{n}, Poland
}%

\date{\today}

\begin{abstract}
The main objective of this work is to investigate theoretically how
tilting of an easy axis of a single-molecule magnet (SMM) from the
orientation collinear with magnetic moments of the leads affects the
switching process induced by current flowing through the system. To
do this we consider a model system that consists of a SMM embedded
in the nonmagnetic barrier of a magnetic tunnel junction. The
anisotropy axis of the SMM forms an arbitrary angle with magnetic
moments of the leads (the latter ones are assumed to be collinear).
The reversal of the SMM's spin takes place due to exchange
interaction between the molecule and electrons tunneling through the
barrier. The current flowing through the system as well as the
average $z$-component of the SMM's spin are calculated in the
second-order perturbation description (Fermi golden rule).
\end{abstract}

\pacs{72.25.-b, 75.60.Jk, 75.50.Xx}


\maketitle

\section{Introduction}
Single-molecule magnets (SMMs)~\cite{Gatteschi_AngewChem42/03} draw
attention as potential candidates for devices which can combine
conventional electronics with
spintronics~\cite{Joachim_Nature408/00}. Characterized by a
relatively large energy barrier for the molecule's spin reversal, a
SMM can be used at low temperatures as a molecular memory
cell~\cite{Timm_PRB73/06}. For these reasons, transport through SMMs
is of current interest~\cite{Timm_PRB73/06, Kim_PRL92/04,
Elste_PRB73/06}. It has been shown that the molecule's spin can be
reversed by a spin current (also in the absence of external magnetic
field)~\cite{Misiorny_PRB07,Elste_CM/0611108}. The phenomenon of
current-induced spin switching is of great importance for future
applications. Furthermore, it is now possible to investigate
experimentally transport through a single
molecule~\cite{Heersche_PRL96/06,Jo_NanoLett6/06,Henderson_CM/0703013}.
However, using present-day experimental techniques, one can hardly
control the orientation of the molecule's easy
axis~\cite{Timm_CM/0702220}.

The main objective of this paper is to investigate theoretically how
tilting of the easy axis of a SMM from the orientation collinear
with magnetic moments of the electrodes (leads) affects the
switching process and current flowing through the system.

\section{model}

The system under consideration consists of a SMM embedded in a
nonmagnetic barrier between two ferromagnetic electrodes. Electrons
tunneling through the barrier can interact via exchange coupling
with the SMM, which may result in magnetic switching of the
molecule. Furthermore, we assume that the spin number  of the SMM
does not change when current flows through the system, i.e. the
charge state of the molecule is fixed. In the case considered here,
the anisotropy axis of the molecule (used as the global quantization
axis $z$) can form an arbitrary angle $\phi$  with magnetic moments
of the leads. To simplify the following description, we neglect the
influence of exchange interaction with the leads on the ground state
of the molecule. Such an influence, however, can be included
\emph{via} an effective exchange field.

\begin{figure}
\includegraphics[width=0.25\columnwidth]{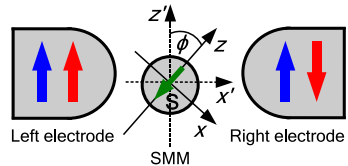}
\caption{\label{fig1} (color online) Schematic picture of the system
under consideration for two collinear configurations of the leads'
magnetic moments: parallel (blue arrows) and antiparallel (red
arrows). The axis $y$ ($y'$) is normal to the $xz$ ($x'z'$) plane. }
\end{figure}

The full Hamiltonian of the system reads
    \begin{equation}\label{eq:FullHam}
    \mathcal{H}=\mathcal{H}_{S\!M\!M}+\mathcal{H}_L+\mathcal{H}_R+\mathcal{H}_T.
    \end{equation}
The first term describes the free SMM and takes the form
    \begin{equation}
    \mathcal{H}_{S\!M\!M}=-DS_z^2,
    \end{equation}
where $S_z$ is the $z$ component of the spin operator,  and $D$  is
the uniaxial anisotropy constant. The next two terms of the
Hamiltonian $\mathcal{H}$ correspond to the two ferromagnetic
electrodes,
    \begin{equation}
    \mathcal{H}_q=\sum_{{\bm k}\alpha} \epsilon_{{\bf
    k}\alpha}^q\: a_{{\bf k}\alpha}^{q\dag}
    a_{{\bf k}\alpha}^q
    \end{equation}
for $q=L$ (left) and $q=R$ (right). The electrodes are represented
by a band of non-interacting electrons with the energy dispersion
$\epsilon_{{\bf k}\alpha}$, where $\bf k$ denotes a wave vector,
$\alpha$ is the electron spin index ($\alpha=+$ for spin majority
and $\alpha=-$ for spin minority electrons), and $a_{{\bf
k}\alpha}^{q\dag}$ ($a_{{\bf k}\alpha}^{q}$) is the relevant
creation (anihilation) operator. Finally, the last term of the
Hamiltonian $\mathcal{H}$ stands for the tunneling processes and is
given by the Appelbaum Hamiltonian~\cite{Appelbaum_PR17/66} rotated
by the angle $\phi$ around the axis $y'=y$ (see Fig.~\ref{fig1}),
    \begin{align}\label{eq:HamTunel}
    \mathcal{H}_T&=\frac{1}{2}\sum_{qq'}\sum_{{\bf k}{\bf
    k}'\alpha\beta}\frac{J_{q,q'}}{\sqrt{N_q\,N_{q'}}}\:
    \bm{\sigma}_{\alpha\beta}\cdot\mathbf{S}\ \Bigg\{\cos^2{\frac{\phi}{2}}\ a_{{\bf k}\alpha}^{q\dag}
    a_{{\bf k}'\beta}^{q'}\: +\: \sin^2{\frac{\phi}{2}}\ a_{{\bf k}\bar{\alpha}}^{q\dag}
    a_{{\bf k}'\bar{\beta}}^{q'}\: -\: \frac{1}{2}\sin{\phi}\Big(a_{{\bf k}\alpha}^{q\dag}
    a_{{\bf k}'\bar{\beta}}^{q'} + a_{{\bf k}\bar{\alpha}}^{q\dag}
    a_{{\bf k}'\beta}^{q'}\Big)
    \Bigg\}
    \nonumber\\
    &+\sum_{{\bf k}{\bf k}'\alpha}\frac{T_d}{\sqrt{N_L\,N_R}}\:
    a_{{\bf k}\alpha}^{L\dag} a_{{\bf k}'\alpha}^R\: +\: H.c.,
    \end{align}
where $\bm{\sigma}=(\sigma^x,\sigma^y,\sigma^z)$  are the Pauli
matrices, $\bm{S}$ is the SMM's spin operator, and
$\bar{\alpha}=-\alpha$. The first term in the above equation
describes exchange interaction of the SMM and electrons in the
leads, with   $J_{q,q'}$ denoting the relevant exchange parameter.
For the sake of simplicity, we consider only symmetrical situation,
where $J_{L,L}=J_{R,R}=J_{L,R}=J_{R,L}\equiv J$. The second term in
Eq.~(\ref{eq:HamTunel}) represents direct tunneling between the
leads, with $T_d$ denoting the corresponding tunneling parameter. We
also assume that $J$ and $T_d$ are independent of energy and
polarization of the leads. Finally, $N_q$ ($q=L,R$)  denote the
number of elementary cells in the $q$-th electrode.

\section{Theoretical description}

The electric current $I$ flowing through the system is determined
with the use of the Fermi golden
rule~\cite{Kim_PRL92/04,Misiorny_PRB07}. Up to the leading terms
with respect to the coupling constants $J$  and $T_d$, the current
can be expressed by the formula
\begin{widetext}
    \begin{align}\label{eq:Current}
    I&=\frac{2\pi}{\hbar}\: e^2\: \Big[|T_d|^2+|J|^2\cos^2{\phi}\left\langle
    S_z^2\right\rangle\Big]\left(D_\uparrow^L D_\uparrow^R +D_\downarrow^L D_\downarrow^R
    \right)\: V
    \nonumber\\
    &+ \frac{2\pi}{\hbar}\: e\: |J|^2\: \sum_m P_m
    \sum_{\eta=+,-}A_\eta(m)\Bigg\{\eta\left[D_\uparrow^L D_\downarrow^R
    \cos^4{\frac{\phi}{2}}+D_\downarrow^L D_\uparrow^R
    \sin^4{\frac{\phi}{2}}\right]\zeta\big(D(\eta2m+1)+\eta eV\big)
    \nonumber\\
    &\hspace{4.9cm}-\eta\left[D_\uparrow^L D_\downarrow^R
    \sin^4{\frac{\phi}{2}}+D_\downarrow^L D_\uparrow^R
    \cos^4{\frac{\phi}{2}}\right]\zeta\big(D(\eta2m+1)-\eta eV\big)
    \nonumber\\
    &\hspace{4.5cm}+\frac{\sin^2{\phi}}{4}\left(D_\uparrow^L D_\uparrow^R +D_\downarrow^L D_\downarrow^R
    \right)\left[\zeta\big(D(\eta2m+1)+\eta eV\big)+\zeta\big(D(\eta2m+1)-\eta
    eV\big)\right]\Bigg\},
    \end{align}
\end{widetext}
where $e$  is the electron charge (for simplicity assumed $e>0$, so
the current is positive for electrons tunnelling from the left to
right). In the above equation  $D_\sigma^q$ is the density of states
(DOS) at the Fermi level in the $q$-th electrode for spin $\sigma$,
and $\left\langle S_z^2\right\rangle=\sum_m m^2 P_m$, where $P_m$
denotes the probability of finding the SMM in the spin state
$|m\rangle$. The voltage $V$ is defined as the difference of the
leads' electrochemical potentials,  $eV=\mu_L-\mu_R$. Finally, we
introduced the notation: $A_\pm (m)=S(S+1)-m(m\pm1)$, and
$\zeta(\epsilon)=\epsilon\big[1-\exp(-\epsilon\beta)\big]^{-1}$ with
$\beta^{-1}=k_B T$.

To compute numerically the current $I$ from Eq.~(\ref{eq:Current}),
one needs to know the probabilities $P_m$. To determine them, the
SMM's spin is assumed to be saturated in the initial state
$|-S\rangle$, and then voltage growing linearly in time is
applied~\cite{Misiorny_PRB07}. Since the reversal process occurs
through all the consecutive intermediate spin states, the
probabilities   can be found by solving the set of relevant master
equations,
    \begin{equation}\label{eq:MasterEquation}
    \left\{
    \begin{aligned}
    c\, \frac{dP_m}{dV}\: &=\: \sum_{l=-S}^{S}\Big\{\gamma_{\: l}^-\:
    \delta_{\: l,m+1}\: +\: \gamma_{\: l}^+\: \delta_{\: l,m-1}\: -\: \left(\gamma_{\: l}^-\:  +\: \gamma_{\: l}^+\right)\: \delta_{\:
    l,m}\Big\} P_{\: l},
    \\
    c\, \frac{dP_{\pm S}}{dV}\: &=\: -\: \gamma_{\pm S}^\mp\: P_{\pm S}\:
    +\: \gamma_{\pm S\mp 1}^\pm\: P_{\pm S\mp 1},
    \end{aligned}
    \right.
    \end{equation}
for $-S<m<S$, where $c=V/t$ is the speed at which the voltage is
increased. The parameters $\gamma_m^{+(-)}$ describe the rates at
which the spin $z$ component ($m$) is increased (decreased) by one.
These tunneling rates have been calculated from the Fermi golden
rule and have the form,
\begin{widetext}
    \begin{align}\label{eq:TransitionTimes}
    \gamma_m^\pm&=\frac{2\pi}{\hbar}\: |J|^2\: A_\pm(m)\: \Bigg\{
    \cos^4{\frac{\phi}{2}}\Big[D_\uparrow^L D_\downarrow^R\: \zeta\big(D(\pm2m+1)\pm
    eV\big)\: +\: D_\downarrow^L D_\uparrow^R\: \zeta\big(D(\pm2m+1)\mp
    eV\big)\Big]
    \nonumber\\
    &\hspace{2.73cm}+\: \sin^4{\frac{\phi}{2}}\Big[D_\uparrow^L D_\downarrow^R\:
    \zeta\big(D(\pm2m+1)\mp
    eV\big)\: +\: D_\downarrow^L D_\uparrow^R\:
    \zeta\big(D(\pm2m+1)\pm
    eV\big)\Big]
    \nonumber\\
    &\hspace{2.73cm}+\: \frac{\sin^2\phi}{4}\Big(D_\uparrow^L D_\uparrow^R\: +\: D_\downarrow^L
    D_\downarrow^R\Big)\: \Big[ \zeta\big(D(\pm2m+1)+
    eV\big)\: +\:  \zeta\big(D(\pm2m+1)-
    eV\big)\Big]
    \nonumber\\
    &+\: \left[\left(\sin^4{\frac{\phi}{2}} +
    \cos^4{\frac{\phi}{2}}\right)\left(D_\uparrow^L D_\downarrow^L
    + D_\uparrow^R D_\downarrow^R \right)\: +\:
    \frac{\sin^2\phi}{4}\left(
    {D_\uparrow^L}^2+{D_\downarrow^L}^2+{D_\uparrow^R}^2+{D_\downarrow^R}^2
    \right)
    \right]\zeta\big(D(\pm2m+1)\big)\Bigg\}.
    \end{align}
\end{widetext}

\section{Numerical results and discussion}

Numerical results have been obtained for the molecule
$\textrm{Fe}_8$~\cite{Gatteschi_AngewChem42/03,Wernsdorfer_Science284/99}
corresponding to the total spin  $S=10$, whose anisotropy constant
is $D=0.292$ K. Apart from this, we assume $J\approx T_d \approx
100$ meV. Calculations have been performed for the temperature
$T=0.01$ K, which is below the molecule's blocking temperature
$T_B=0.36$ K, and for $c=10$ kV/s. It has been also assumed that
both the leads are made of the same metallic material characterized
by the total DOS $D=D_+^q+D_-^q\approx 0.5$ per electron-volt and
per elementary cell, where $D_{+(-)}^q$ denotes the DOS of majority
(minority) electrons in the $q$-th electrode. Furthermore, the
$q$-th electrode is described by the polarization parameter $P^q$
defined as $P^q=(D_+^q-D_-^q)/(D_+^q+D_-^q)$. The following
discussion is limited to the case, where one electrode (the left
one) is fully polarized, $P^L=1$, whereas the polarization factor of
the second electrode can vary from $P^R=0$ (nonmagnetic) to $P^R=1$
(half-metallic ferromagnet).

\begin{figure}
\includegraphics[width=0.25\columnwidth]{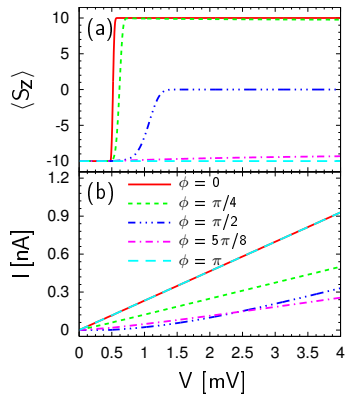}
\caption{\label{fig2} (color online) The average value of the SMM's
spin, $\langle S_z\rangle$, and the current  $I$  flowing through
the system as a function of the voltage $V$, calculated for the
parallel configuration of the electrodes' magnetic moments with
$P^L=1$ and $P^R=0.3$. }
\end{figure}

In Fig. 2 we show the average value of the $z$-component of the
SMM's spin, $\langle S_z\rangle$, and the charge current $I$,
calculated for several values of the angle $\phi$   and for parallel
magnetic configuration of the leads. The case of  $\phi=0$
($\phi=\pi$) corresponds to the situation when the initial SMM's
spin is antiparallel (parallel) to the leads' spin moments. One can
note that the influence of current on the molecule's spin gradually
disappears as the angle $\phi$ approaches $\pi$. For  $\phi<\pi/2$,
the molecule's spin becomes switched from the state $|-S\rangle$ to
the state $|S\rangle$. The switching time, however, becomes longer
and longer as the angle $\phi$ approaches $\phi=\pi/2$.  At
$\phi=\pi/2$, which corresponds to the situation with the SMM's easy
axis perpendicular to the leads' magnetic moments, different
molecular spin states $|m\rangle$ become equally probable with
increasing voltage, and therefore  $\langle S_z\rangle\rightarrow
0$. This is a consequence of the fact that when the voltage exceeds
the activation energy for the spin-flip
process~\cite{Misiorny_PRB07}, the SMM undergoes transitions to
upper and lower spin states with equal rates $\gamma_m^+=\gamma_m^-$
(see Eq.~(\ref{eq:TransitionTimes})).

When  $\phi>\pi/2$, the spin state of the molecule is only weakly
modified by current, and remains strictly unchanged for $\phi=\pi$.
The absence of switching by positive current at large values of
$\phi$ (for the assumed parameters) is consistent with the
conclusion of Ref.~[6], where for collinear configurations and
positive current only switching from $|-S\rangle$ to $|S\rangle$
states was allowed, whereas positive current had no influence on the
state $|S\rangle$.

The SMM's spin can be reversed due to exchange interaction with
tunneling electrons. The latter flip their spins and hence add to or
subtract some amount of angular momentum from the molecule. As the
angle $\phi$  grows, the spin orientation of tunneling electrons
`seen' by the molecule and consequently also the transition rates
given by Eq.~(\ref{eq:TransitionTimes}) change as well.
Figure~\ref{fig2}b shows the current flowing in the system as a
function of the bias voltage. This current strongly depends on the
orientation of the SMM's easy axis. This dependence is a consequence
of the fact that the dominant contribution to current is due to the
exchange term (first term in Eq.~(\ref{eq:HamTunel})), which is
sensitive to the orientation of the SMM's spin. The curves for
$\phi=0$ and $\phi=\pi$ overlap (except for a small voltage range
where the switching for $\phi=0$ takes place -- not resolved in
Fig.~\ref{fig2}b).

\begin{figure}
\includegraphics[width=0.5\columnwidth]{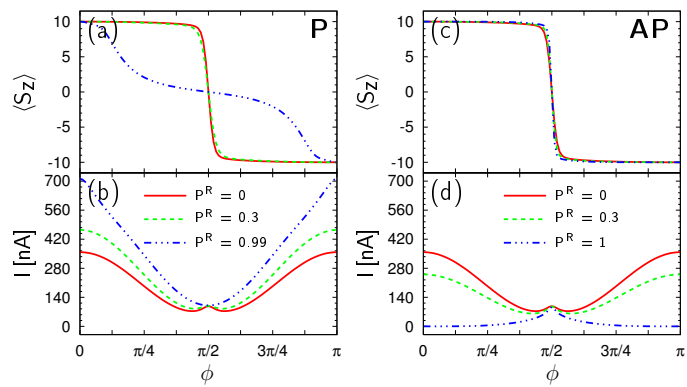}
\caption{\label{fig3} (color online) The average value of the SMM's
spin $\langle S_z\rangle$  (a), and the current  $I$  flowing
through the system (b) as a function of the angle $\phi$  in the
parallel (P) (a,b) and antiparallel (AP) (c,d) magnetic
configurations for $V=2$ mV and  $P^L=1$.}
\end{figure}

Figure~\ref{fig3} presents the spin $z$-component $\langle
S_z\rangle$ and the current $I$ in both configurations of the leads'
magnetic moments, plotted as a function of the angle $\phi$  and
calculated for $V=2$ mV. For  $\phi<\pi/2$, the spin switching takes
place in both parallel and antiparallel magnetic configurations.
Figure~\ref{fig3} also indicates that the current at $\phi=\pi/2$ is
independent of the magnetic configuration as well as on the
polarization parameters of the leads.

In conclusion, we have shown that tilting the easy axis of a SMM
from the collinear orientation relative to the leads' magnetic
moments has a significant influence on the reversal process of the
molecule's spin, as well as on current flowing through the system.

{\it Acknowledgements} This work was supported by funds from the
Ministry of Science and Higher Education as a research project in
years 2006-2009. One of us (MM) also acknowledges support from the
MAGELMAT network.

\end{document}